**Title: The Moral-IT Deck: A Tool for Ethics by Design.**

Authors: Lachlan Urquhart and Peter Craigon.

Biographies:

Dr Lachlan Urquhart, Lecturer in Technology Law, Department of Law, Old College, South Bridge, University of Edinburgh, Edinburgh, EH8 9TY
Corresponding author: lachlan.urquhart@ed.ac.uk

Dr Peter Craigon, Research Fellow, Future Food Beacon of Excellence, University of Nottingham, Sutton Bonington Campus LE12 5RD
Horizon, University of Nottingham.

# The Moral-IT Deck: A Tool for Ethics by Design

**Abstract:**
This paper presents the design process and empirical evaluation of a new tool for enabling ethics by design: The Moral-IT Cards. Better tools are needed to support the role of technologists in addressing ethical issues during system design. These physical cards support reflection by technologists on normative aspects of technology development, specifically on emerging risks, appropriate safeguards and challenges of implementing these in the system. We discuss how the cards were developed and tested within 5 workshops with 20 participants from both research and commercial settings. We consider the role of technologists in ethics from different EU/UK policymaking initiatives and disciplinary perspectives (i.e. Science and Technology Studies (STS), IT Law, Human Computer Interaction (HCI), Computer/Engineering Ethics). We then examine existing ethics by design tools, and other cards based tools before arguing why cards can be a useful medium for addressing complex ethical issues. We present the development process for the Moral-IT cards, document key features of our card design, background on the content, the impact assessment board process for using them and how this was formulated. We discuss our study design and methodology before examining key findings which are clustered around three overarching themes. These are: the value of our cards as a tool, their impact on the technology design process and how they structure ethical reflection practices. We conclude with key lessons and concepts such as how they level the playing field for debate; enable ethical clustering, sorting and comparison; provide appropriate anchors for discussion and highlighted the intertwined nature of ethics.

**Acknowledgements:** Funded by EPSRC Grant EP/M02315X/1. We also want to thank Dr Dimitri Darzentas for his contribution to the project.

**Introduction.**

Building ethical systems can be difficult and we believe better tools are needed to support the role of technologists in addressing ethical issues during system design. Accordingly, we designed, tested and evaluated a new approach to enable ethics by design: **The Moral-IT Cards**. They support reflection by technologists on normative aspects of technology development, particularly at the early stages. They provide *anchoring* content and processes to structure reflection on emerging risks, appropriate safeguards and challenges of implementing these in the system.

In this paper, we document how these cards have been developed and tested within 5 workshops with 20 participants from both research and commercial settings. The cards pose questions, requiring critical reflection and situated judgements on how best to proceed with value judgements. They were used with the *Moral-IT Impact Assessment Board* to structure and manage deliberation on ethical responsibilities.

We will begin by mapping out the field in relation to cards and ethics. We do this by considering the turn to the role of technologists in regulation, particularly how the normative dimensions of their role sit alongside functional dimensions. This shift highlights the need for practical support to enable creators to consider, engage with and address their responsibilities, thus we consider existing mechanisms for bringing ethical reflection into design. We then unpack the value of our tool and why cards are a useful medium for reflection on complex issues.

We then introduce the content of the Moral-IT cards, documenting key features of our card design, background on the content, and the process for using them. This involves unpacking the sources that informed our impact assessment process Board, before discussing our methodology for the workshops, use cases and our key findings. These are clustered around three overarching themes: the value of our cards as a tool, their impact on the technology design and how they structure ethical reflection practices. Lastly, we provide a series of reflections on how card based tools can be used and the value of our tool for doing ethics by design.

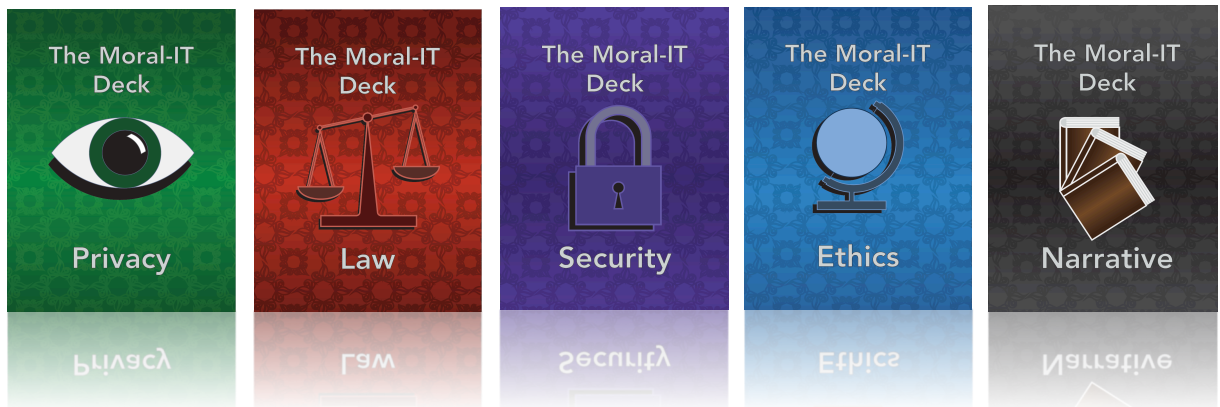

**Part I – Context and Motivation.**

In designing our cards, we were influenced by how different academic disciplines have recognised the role of technologists in dealing with normative aspects of their work (particularly legal and ethical issues) (Urquhart, 2018). Below we document key insights that underpin the role of technologists in ethics by design and set context for why design tools like cards are needed. We draw from Science and Technology Studies (STS), IT Law, Human Computer Interaction (HCI), Computer/Engineering Ethics, contemporary policymaking (EU and UK focused) and one which spans each of those above, Responsible Research & Innovation (RRI). For clarity, we consider each in turn.

*STS:* In the 1970s, prominent STS scholars such as Winner (1978) recognised that the design of technology has normative impacts, stating "*technology in a true sense is legislation. It recognizes that technical forms do, to a large extent, shape the basic pattern and content of human activity in our time*" (p323). Similarly, Latour (1992) argued technology design involves normative decision making that delegates power to non-human entities to permit or prohibit user behaviours, as did Akrich (1992). As Latour states: "*the distance between morality and force is not as wide as moralists expect, or more exactly, clever engineers have made it smaller*" (p174). Clearly, the role of technologists in ethics is a longstanding concern.

*IT Law:* In legal scholarship, scholars such as Reidenberg (1998), Lessig (1999), Murray (2008), Leenes (2011) and Hildebrandt (2016) have similarly debated how technology design can shape human behaviour. Here, embedding legal values in technology design enables enforcement of legal norms in a domain where jurisdictional and practical issues of fast technological changes have challenged traditional institutions of the law (e.g. courts, legislatures). Some have questioned the legitimacy of non-state actors (like technology companies) in controlling citizens behaviour through system design (Leenes, 2011). The nature of checks and balances to safeguard due process and the 'rule of law' are particular concerns, particularly around retaining negotiability and interpretation of legal rules (Brownsword, 2008; Hildebrandt, 2016). For ethics, we can learn from and address these concerns around legitimacy of technologists. Urquhart (2018) argues HCI specialists could have some regulatory legitimacy due to their proximity to users and methodological tools (e.g. participatory or co-design) to understand and respond to their needs in more granular ways. We now turn to how HCI researchers and practitioners position themselves in this domain.

*HCI:* There are longstanding concerns around how the HCI community practically engages with normative issues in design (Shneiderman 1990). For example, Flanagan, Howe and Nissenbaum (2008) note there are challenges of terminology and methodologies where "*even conscientious designers, by which we mean those who support the principle of integrating values into systems and devices, will not find it easy to apply standard design methodologies, honed for the purpose of meeting functional requirements, to the unfamiliar turf of values*" (p323). Lazar et al (2012) mirrors these concerns. Value sensitive design and reflective design have been strong practical influences for our cards in finding practical routes forward to support technologists. With the latter, designers need to consider their role and impacts on users to "*bring unconscious aspects of experience to conscious awareness, thereby making them available for conscious choice*" by highlighting and questioning assumptions, ideologies, and beliefs of design (Sengers et al, 2005, p50). This includes considering reflection on their position, knowledge and impact on users (Grimpe et al, 2014). We build from Nissenbaum's (2001) point that "*systems and devices will embody values whether or not we intend or want them to. Ignoring values risks surrendering the determination of this important dimension to chance or some other force*" (p119). Our cards seek to support the process of reflection by asking probing questions (prompting answers, and action). To better understand the importance of technologists' reasoning, we turn to Computer Ethics.

*Computer Ethics*: Prominent technology ethicist Verbeek (2006) has stated "*engineering design is an inherently moral activity*" (p368) and similarly, Millar (2008) states "*in effect, engineers ought to be considered de facto policymakers, a role that carries implicit ethical duties*" (p4). A sense of 'moral overload' is a risk for technologists (Van den Hoven et al, 2011) and our cards aim to provide a means of working through difficult, practical moral decisions involving value trade-offs and balancing competing interests during design. Other ethical approaches such as anticipatory ethics (Brey 2012) and the ETICA approach (Stahl 2011), have emerged to insert technology ethics thinking into value judgements during design. This often entails social scientists collaborating with, assessing and engaging with scientific processes and stakeholders, which can be time consuming, constrained by setting and conflict driven (Fischer, Mahajan and Mitcham 2006; Felt, Fochler and Sigl 2018 p205). We consider these further below, but first conclude by considering how Responsible Research and Innovation (RRI) informed our cards.

*RRI*: This domain focuses on ethical, legal and social implications (ELSI) of innovation, (Zwart et al 2014) considering how to integrate RRI approaches into researchers' practices, as opposed to this being someone else's problem (Felt, Fochler and Sigl, 2018 p202). Driven by EU policymakers and funding councils (Burget et al 2017), it encourages scientists and innovators to take "*care of the future through collective stewardship of science and innovation in the present*" (Stilgoe et al 2013 p1570). It unpacks responsibility, calls into question who is responsible, for what and reflects on wider impacts of innovation (Von Schomberg, 2013). RRI has different framings but fundamentally requires four key steps of anticipation, reflection, engagement and action (EPSRC 2020). The cards support RRI, insofar as they seek to encourage anticipation of issues, encourage reflection on the implications, prompt responses, and mandate action (in answering the questions).

***Policymaking in EU and UK***: Alongside this theoretical backdrop we also see policymaking shifts around enacting design led solutions to normative problems. These include Articles 25 and 32 of the EU General Data Protection Regulation 2018 requiring those collecting personal data to utilise organisational and technical safeguards to support 'data protection by design and default' and good security management, respectively.(European Data Protection Board 2019) Similarly, the UK government Department of Digital, Culture Media and Sport (DCMS) report on 'Secure by Design' for the IoT framework provides recommendations to industry for building more secure embedded, networked systems.(DCMS 2018)  Another example is the UK Information Commissioner's code of conduct for age appropriate design (ICO 2020) which focuses on impacts of defaults and technical design choices on child rights.

Whether design led policy approaches work effectively remains to be seen, for example realising privacy by design in practice remains a complex, cross-disciplinary exercise with mixed results (Danezis et al 2015). Similarly, the 2019 UK White Paper on Online Harms proposes creation of a 'Safety by Design' framework to technically address dissemination of hate speech, terrorist propaganda, online misogyny and fake news (DCMS 2019). Responses to a recent UK consultation showed significant confusion around what this actually requires, especially around practical steps to respond to safety responsibilities (DCMS 2020). Similarly, 'technical protection measures' mandated by international copyright law have long been controversial mechanism for enforcing copyright law norms (European Parliament 2001).

Nevertheless, we see an academic and policy direction that implicates technologists in dealing with normative harms and risks. Thus, we now consider existing 'ethical design tools' before considering what card-based approaches might offer, and then introduce our Moral-IT cards' content and process for using them.

***Existing Ethical Design Tools****.*

There have been numerous attempts to support virtuous behaviour by technologists (Volkman, 2018) with work from the ACM, British Computer Society (BCS 2020), IEEE and many other industry, third sector and governmental stakeholders (Field and Nagy 2019). By no means exhaustively, Table 1 below covers our list of example tools seeking to support reflection by technologists on their work. Impact assessments and ideation cards, which we draw on more extensively are discussed below.

*Table 1 Ethics by Design Tools*

| **Tool** | **Source** |
| --- | --- |
|  |  |
| **Ethical codes of practice** | Belmont Principles for Ethics (Belmont Report 1979), <br> ACM/IEEE Codes of Practice (ACM 2018, IEEE nd, Gotterbarn 1991); <br> Royal Academy of Engineering/Engineering Council (Royal Academy of Engineering. nd); <br> IEEE 'Ethically Aligned Design' initiative (IEEE ethics in action. nd) |

| Ethical technical standards e.g IEEE P7000 Standards | (IEEE, 2018) |
|---|---|
| **Ethical matrix** | (Mepham et al 2006, Forsberg 2007) |
| **Ethical checklists** | (Verharan and Tharakan 2010; McStay and Pavlisack, 2019) |
| **Scenario-based tools** | (Ikonen et al 2012) |
| **RRI search engine, training and self-reflection tools** | (Groves, 2017). |
| **Ethical stack** | (Virt.EU, 2020) |
| **Data ethics canvas** | (ODI, 2019) |
| **Ethical games** | (Belman et al, 2011) |

*Why Cards as a Medium?*

In part, these cards build on experiences with card-based tools, particularly for data protection governance (Luger, Urquhart, Rodden and Golembewski, 2015*)*. However, card-based tools have a history in design and computing going back to early uses by Neilsen (1995) and Muller (2001). Table 2 provides our illustrative, but non-exhaustive, list of card-based tools which are often given different names, e.g. ideation (Selby and Golembewski 2010), method (IDEO 2003), pattern (Wetzel, 2011), envisioning (Friedman and Hendry 2012) cards. The Moral-IT deck, as described here, function primarily as *evaluation* cards, where the goal is to support structured reflection on design choices[1]. Part of their value is translating ethical knowledge into an accessible, tangible form to enable technologists to question their practice in a more playful way.

*Table 2 Example Decks of Cards*

| Application domain for the deck | Sources |
|---|---|
| **Creativity in design.** | Golembewski and Selby (2010) |
| **Value sensitive design.** | Friedman and Hendry (2012) |
| **Use of methods in user experience testing.** | IDEO (2003) |
| **Mixed reality game design.** | Wetzel et al (2014). |
| **Computer security threats.** | Denning et al (2013) |
| **Computer security education.** | Denning et al (2013b) |
| **Addressing algorithmic fairness in information systems.** | Lane, Angus and Murdoch (2019) |
| **Gig economy service design.** | Fedosov et al (2019) |
| **Responsibility in research** | Felt, Fochler and Sigl (2018) |
| **Information privacy.** | Luger, Urquhart, Rodden and Golembewski (2015); Urquhart (2016); Barnard Wills (2012). |
| **Human values in games.** | Belman et al (2011) |
| **Sustainable Economy** | Innovate UK (2016) |
| **Policy design methods.** | SILK (2007). |
| **Addressing creative blocks.** | Eno and Schmidt (1975) – these inspired our Ace cards. |
| **Internet of Things device design.** | Know Cards (2016) |

---

[1] The cards have been designed to be flexible and adaptable in their use so can be used in alternative ways to the evaluative approach described here. Further work will explore this potential in more detail.

| **Structuring Debate of Complex Issues** | PlayDecide (2018) |
|---|---|
| **Collaborative design of software requirements.** | Tudor, Muller, Dayton & Root (1993) |
| **Playful experiences** | Lucero and Arrasvuori (2010) |

In unpacking the value of cards further, Roy and Warren (2018) reviewed the use of cards in design research and found that they had been used to serve a variety of purposes. Focusing on the subject matter and content of the cards, this includes (p137):

- 'creative thinking and problem solving'
- 'systematic design methods'
- 'human centred design'
- 'domain specific design'
- 'futures thinking'
- 'collaborative working'

Felt, Fochler and Sigl (2018) think more about how cards function and, state their IMAGINE RRI cards provide a 'narrative infrastructure' as a tool to help discuss issue around RRI (p203) i.e. to "*stimulate researchers' capacity to reflect on the social and ethical aspects of their work and can be applied and adapted relatively widely with limited effort.*" (p205). Similarly, Luger, Urquhart, Golembewski and Rodden (2015) found their 'privacy by design ideation cards' helped designers engage with law and challenge the idea that law is something remote to their role and only for specialists. Instead, the cards showed designers have a regulatory role and such tools can support engagement with that in a creative manner during the design process.

Roy and Warren, (2018,) document the key strengths of card-based tools as providing clear information delivery and enabling communication. From other studies (e.g. Carneiro, Barros and Costa, 2012; Casais, Mugge & Desmet 2016) show this is enabled through the summarised format for information presentation, the physical nature of cards and their ability to enable bringing together ideas in novel ways. Cards can structure discussion (Sutton 2011; Hornecker, 2010), whilst also playing a role in unpacking issues through channelling attention and enabling tangible uses such ranking/ordering/classifying (Kitzinger, 1994). Card based tools are not perfect however, with issues including; providing 'too much' or 'oversimplified' information, being difficult for users to use and the lack of scope to change/update (Roy and Warren, 2018, p131).

Turning further to literature in this domain we see longstanding assertions that playful, game like tools such as cards are valuable as '*having objects at hand to play with is important as it speeds up the process and help participants to focus. As design material game pieces and props create a common ground that everybody can relate to and at the same time they act as 'things-to-think-with'*'(Papert 1980, Brandt and Messeter, 2004 p129). Cards also act as physical anchors for discussion where they "*afford actions such as pointing, grabbing, grouping, and sorting. Cards support participants in externalizing design rationale and analysis, thus making ideas more concrete and accessible to themselves and to their partners*" (Deng et al, 2014 p8). We draw on and further develop this notion of cards as anchors throughout this article. By extension the opportunity to provide users with a 'tool to think with' and externalise discussions around ethics points to the potential value of a card based tool in the development process.

Whilst literature points to the novelty and utility of card-based tools, there is little consideration paid to what cards *are* as a medium, with more attention being paid to their content. One exception is Altice (2016) who surveys cards in games to identify five common characteristics that make up playing cards as a 'platform'. He contends that cards are:

- *Planar* - they are two dimensional and lay flat,
- *Uniform* - in size and shape,
- *Ordinal* - in terms of having order and ranking between card,
- *Spatial* - where the layout and interaction with them in space is significant
- *Textural* – Optimised for handling and touch and 'hands' to interact with and enable other characteristics.

Much of the appeal of card-based tools appears to draw from the familiarity with these aspects, which in turn are drawn from card games. These characteristics suggest why cards differ from simple pieces of paper or information delivered and manipulated in a different media.

To benefit from the value of cards, we created a tangible, material deck as opposed to a mobile application or a website. However, digital modularity would be more easily updatable (especially when dealing with normative frameworks like law and ethics where values shift). In contrast, once text is committed to the cards, and printed, it is harder to change these (although workarounds such as blank cards, stickers or writing on them in pen can work) but adding additional cards at a later date need not affect the integrity of the existing cards. Our cards have been developed to be machine readable through interaction with an augmented reality tool for digitally tracking their use, but they remain a physical artefact (Darzentas et al 2019).

This nature and affordances of cards is an aspect we develop further below, demonstrating the value of cards as anchors. Now we turn to discuss in more detail our creation of the Moral-IT card deck.

***Introducing the Moral-IT Cards***

***Card Content.***

These cards are a design probe (Sharp, Rogers and Preece 2019, Wallace 2013)[2] to provoke discussion around practical ethical questions. In part, they are inspired by Verbeek's argument that ethical decisions are mediated and answered through design practice (Verbeek, 2005). For engineers and designers, ethical dilemmas and resolutions occur at a grounded, practical level (Millar, 2014; Verbeek 2006), and we were interested in observing those deliberations, as opposed to focusing on formulating more abstract, absolute framings of ethical practice (as codes of ethics often do). The cards provide an anchor to consider 'ethical' issues (broadly framed) and think about what 'ought' to be done to design more responsible systems.

Table 3 contains our overview of the Moral-IT deck[3]. We do not claim the ethical groupings or issues on these cards are exhaustive or definitive (if that can ever be claimed) and instead these are starting points for discussion. The cards reflect the authors' multidisciplinary training primarily in computing, technology law, human computer interaction, STS and critical theory. We also provided blank cards within the deck so participants could add their own or flag any missing concepts, in itself helping us test the robustness of our suggestions. For example, whilst the concept of consent is contained within cards such as 'Special Categories of Data', some participants sought to find a specific 'consent' card. Also, discussions raised issues of how a system or data may be used by different parties with concerns about data misuse being raised which may highlight the need for a specific card to cover it. If recurrent issues keep arising then this would indicate that a specific new card may need to be added to the deck. We also have a series of 'narrative' cards which are an alternative mechanism to the Board for using cards. These are valuable when hypothetical scenarios are needed, users are doing the activity independently or looking for a shorter approach e.g. in teaching. The narrative cards require users to construct a technology scenario and assess its risks using cards. We do not provide analysis on them here, focusing instead on 'real life examples' and the process Board, but see appendix 2 for reference to narrative card content.

---

[2] Sharp, Rogers, and Preece, (2019) p399 where they state "design probes are objects whose form relates specifically to a particular question and context. They are intended to gently encourage users to engage with and answer the question in their own context";
[3] See full deck available here too online, as downloadable PDF - https://lachlansresearch.com/the-moral-it-legal-it-decks/

In part, our breadth of issues covered is inspired by Moor's framing of computer ethics as questioning "*the nature and social impact of computer technology and the corresponding formulation and justification of policies for the ethical use of such technology*" (Moor 1985 – p266). Many issues can come under the remit of IT ethics and we wanted to avoid these being framed as an absolute checklist that if followed, then a design is approved as 'ethical'. Instead, the cards sensitise technologists to ethical issues, a tool to encourage critical discussions around ethical responsibilities in design and to help develop a questioning mindset.

*Table 3 List of Moral-IT Cards*

| Card Number | Suit | | | |
|---|---|---|---|---|
| | Security / Diamonds | Ethics / Spades | Privacy / Hearts | Law / Clubs |
| [2] | Identities Management | Legibility and Comprehension | Limited Data Collection | Environmental Protection |
| [3] | Obfuscation | User Empowerment and Negotiability | International Data Transfer | Accessibility |
| [4] | Secrecy | Overt Bias and Prejudice | Spectrum of Control Rights | Consumer Protection |
| [5] | Trustworthiness | Autonomy and Agency | Transparency Rights | Rule of Law |
| [6] | Confidentiality | Trust | Lawful Processing | Due Process |
| [7] | Usable Security | Meaningful Transparency | Data Security | Risk Minimisation |
| [8] | Resilience and Low Redundancy | Sustainability and eWaste | Taking Responsibilities | Liability |
| [9] | Data Breach Management | Power Asymmetry | Privacy in Public | Proportionality |
| [10] | Physical Safety | Fairness and Justice | Location Privacy | Precautionary Principle |
| [Jack] | Attribution and Responsibility | Temporality | Compliance and Accountability | Duty of Care |
| [Queen] | Integrity | Wellbeing | Special Categories of Data | Intellectual Property |
| [King] | Secure for Whom? | Participation | Privacy Virtues | Criminality |
| [Ace] | What's the most embarrassing thing about your technology? Would you change it? How? | Consider the setting this technology will be used in and why this is important0 | Think of a time you were amazed by a new technology. Why? | How can your technology embody human virtues? |

We are also sensitive to criticisms faced by Friedman et al (2008) in their envisioning cards and value sensitive design work. Le Dantec et al (2009) and Borning & Muller (2012) questioned how they formulated their 'values with ethical import' i.e. "*what a person or group of people consider important in life*" (p70). Valid critiques of whose perspectives are around what is being prioritised and why, with one angle stating they focus on Western democratic 'liberal' values such as privacy, justice, autonomy for example. The values in our cards are not to be seen as an exclusive, definitive or exhaustive list of principles. Instead, they are a mechanism to begin a conversation and to then critique if these are indeed the right values. As they use questions, the responses to those acts as a starting point for further reflection on the merits of the value. The open questions can be tested, rejected, replaced or refined by the groups - they are not static or immutable, even if the medium (e.g. cards) make this appear to be the case. As we shall see, there is interpretive flexibility of the cards, borne out in the workshops. Using them as a probe, we were able to test them with different groups and obtain their feedback. We clustered our principles under suits of security, ethics, privacy and law (see table above), and these suits capture many of the legal, ethical and social concepts we deem important when designing (information) technologies. Below we unpack the rationale behind one card, as an example of the thought process which underpins the design and creation of each card in the whole deck.

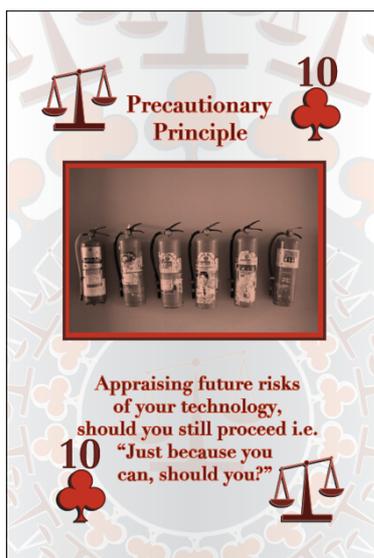

*Figure 1 Example Moral-IT Card -Precautionary Principle, Law Suit*

The **precautionary principle** (Figure 2) is often both legal and ethical best practice for uncertainty around risk from new technology (Fisher 2006). In Europe, 'command and control', state led regulation of emerging technologies often takes precedence. Data protection laws for IT, labelling rules for nanotech and limitations on biotech all shape emergence of new innovations. (European Commission 2017; EDPB, 2019) In environmental law specifically, the precautionary principle is prescriptive around preventing harm from pollution or environmental disaster (UN 1992). This principle also aligns with more anticipatory forms of governance which seek to guard against future and intersects with RRI concepts of stewardship and anticipation, requiring reflection by innovators on future risks.

**Card Style:** We developed a traditional 'playing card' deck style to embrace a game orientated aesthetic[4], but also to constrain the number of substantive cards (52) and to force us to formulate 4 suits to cluster our ethical questions. The card text was designed to be provocative and the use of a direct but open question was intended to ensure that the principle could not be dismissed easily, instead requiring an explanation to be formulated to promote engagement. The openness also enabled conceiving of the card flexibly for example as both a risks and safeguards e.g. a risk of trustworthiness could be dealt with by meaningful transparency. Our questions are focused on 'technology', as opposed to just 'computing' or 'systems', to broaden utility.[5] We designed playful suit markers alongside traditional club, heart, diamonds and spades[6] with a padlock for security; eye for privacy; scales of justice for law; and a globe for ethics. With our 4 Ace cards, inspired by Eno and Schmidt's (1975) 'oblique strategy' card work, we posed some more abstract, provocative questions e.g. 'what is the most embarrassing thing about your technology?'. Images are important in card design, for triggering thought processes or emotional reactions, in addition to aesthetic reasons. (Friedman and Henry 2012) We chose the images to illustrate the principles in a variety of literal or abstract ways, which was intended to provoke questioning and reflection and promote discussion. We also chose images[7] to enable trackability of cards for future compatibility with the Cardographer augmented reality platform (Darzentas et al 2019).

---

[4] In past research with cards, we often found researchers wanted thought they were a game, hence with this deck we adopted this aesthetic.
[5] NB they were designed in a computing context, used in workshops with projects utilising embedded sensors, affect sensing, location tracking, tangible and mobile computing, mixed media repositories etc.

Now we discuss how we formulated our *Moral-IT IA Board* process for using them an example of potential ways of using them that we used for our testing and evaluation as discussed through the remainder of this paper.

*Creating the Moral-IT IA Board*

Cards can be sorted, ordered, clustered, ranked and discarded amongst other actions. Through the development of the cards we sought to keep the potential way(s) of using the cards open and flexible so that users could create and adapt them to their own use to broaden their appeal. In order to test and evaluate the cards in our workshops however we needed to develop a use scenario to test. We focused on creating a streamlined 'impact assessment' to structure discussion. Our *Moral-IT IA Board* identifies 4 key stages pertinent to considerations of 'ethics by design' namely: identifying possible risks; assessing significance or importance of the risk; establishing suitable safeguards to these risks; and exploring practical implementation challenges.

Impact assessments (IAs) for new technologies are a popular tool in forecasting issues and designing strategies for action in a structured manner e.g. a privacy impact assessment is a "*systematic process for evaluating the potential effects on privacy of a project, initiative or proposed system or scheme and finding ways to mitigate or avoid any adverse effect*s" (Wright 2011) Figure 1 shows a variety of IAs we were inspired by in this work.

---

[6] Retaining these enables gameplay with cards as traditional decks.
[7] We used contextually appropriate images from Pixabay (for intellectual property reasons) and matched these to the text.

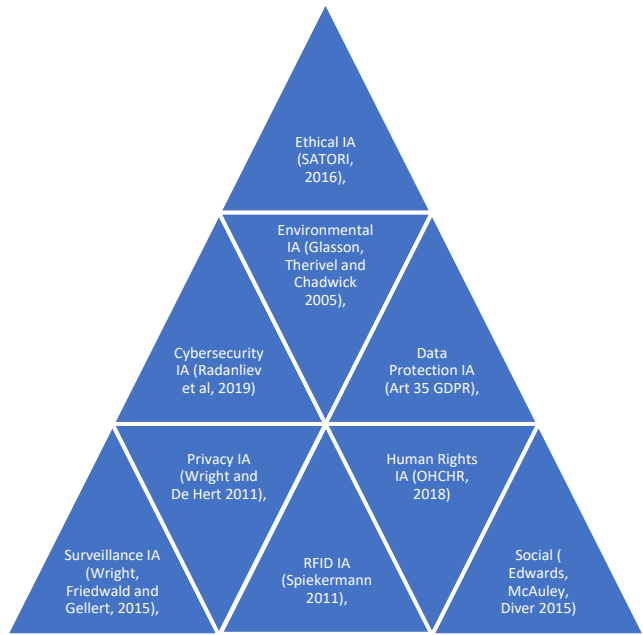

*Figure 2 Example Impact Assessments.*

IAs have uses in a variety of domains and despite benefits, they can be very resource and time intensive (Morrison - Saunders et al 2015). We were particularly concerned that individuals and organisations sometimes lack the resources or motivation to engage with ethical issues in the first place, and thus a more streamlined tool could help. Technology SMEs[8] and start-ups face difficulties in dealing with legal/ethical concerns (Norval et al 2019) and there is value in low-cost card-based tools to support them (Urquhart, 2016; Urquhart 2020). We need to learn from criticisms against older IAs and similar processes e.g. constructive technology assessment (Rip and Te Kulve 2013) that they can be highly time intensive and they do not generate wider solutions as a result (Felt, Fochler and Sigl, 2018). Furthermore, there are concerns from the wider RRI and technology ethics community[9] around the need to capitalise on researcher knowledge (as opposed to just ethicists) (Brey 2000) and to foreground their perspectives in such processes (Le Dantec Poole and Wyche 2009 pg1141, Borning and Muller pg1125 2012, Reijers et al 2018 pg 1455). Reijers et al, for example, state that tools for '*ethical technology design should focus more on the integration of ethics in the day to day work of R and I Practitioners…*' (Reijers et al p1457). Similarly Felt et al, have argued for the need to really include researchers in a collaborative arrangement, not outsourcing to them at convenient times in the project and avoiding a 'new bureaucracy of virtue' with RRI as tick box exercises (Felt, Fochler and Sigl 2018 pg 202) As a lightweight, adaptable tool, the Moral-IT Deck and Board is amenable to integrating with technologists working practices without stopping them 'getting on' with the doing of the research and development. We now turn to how the cards were used in our workshops, before presenting our findings

---

[8] Small and Medium Sized Enterprises
[9] As Felt, Fochler and Sigl (2018) state "*any successful RRI activity must find a way of making RRI a core element in research practice, despite competing values. If this can be achieved, RRI related work potentially play the role of a 'moral glue that holds the often simultaneous yet potentially contradictory promises of economic, societal and scientific benefits together*"

# Part II: Methodology.

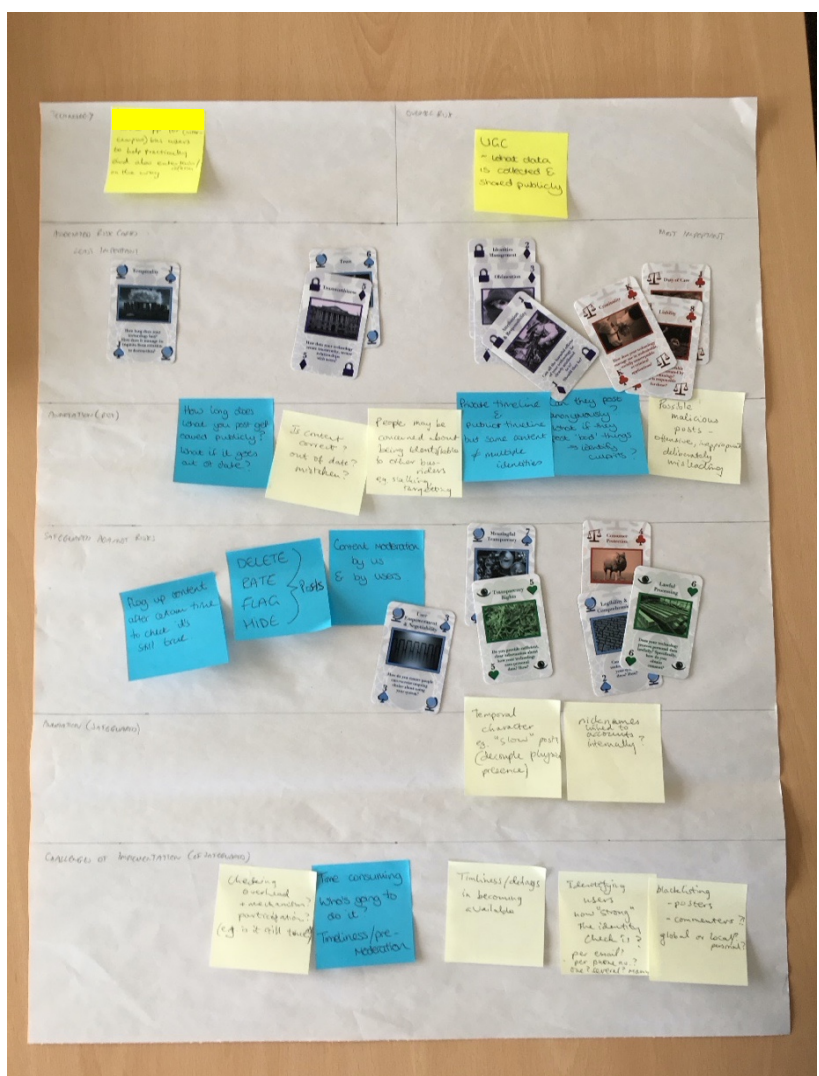

*Figure 3 ST group Completed IA Board*

The cards were tested in a series of workshops with 5 groups, lasting an average of 2.5 hours. 4 groups were working on the development of technology in a research setting; 1 group broadly worked in software and technology development in industry. Group demographics are summarised in table 4 below. Whilst there were pre and post workshop questionnaires conducted, we focus on qualitative data from our groups using cards with the impact assessment board in a systematic manner, with a completed board provided below in figure 3. This provides a useful material artefact, making ethical deliberations visible, and demonstrates some of the card use practices e.g. clustering, ranking[10]. We will present other snapshots of boards in subsequent sections. The cards that were used by each of the groups and their order of ranking can be seen in the appendix. Participants followed this sequence:

- o **Defining the technology** – Summarising what their technology was and writing this on a post-it note at the top left of the physical impact assessment board. This could be a real or hypothetical system (depending on the group).
- o **Defining the main ethical risk** – Providing an overall ethical risk for their technology. Whilst multiple risks exist, choosing an overarching one helped focus discussion). This was written on a post-it note and placed on the top right.
- o **Associated risks** – They were advised to pick their most important 5 cards[11] from the Moral-IT deck that they felt were most associated with their overall ethical risk. The decision-making process was not prescribed and left to the group to decide.

---
[10] Such artefacts would be a useful ongoing record and working document that groups could refer to, update and amend as they progressed through the development of their technology, much like a privacy impact assessment can be an organic document.
[11] This was a guide and not a prescribed figure, some clustered more cards under 5 themes, some picked less as we see in the results.

- o **Ranking** – The groups were then asked to rank and arrange these (5) cards they had selected from least important (on the left) to most important (on the right) and place them in the row provided on the process board. (Figure 3)
- o **Annotating Risks** – Participants were asked to annotate some of the reasoning behind their choice of cards on post-it notes and sticking these directly below the chosen card on the line marked annotations.
- o **Safeguards** – On the next line of the process board the participants were asked to use the cards to identify principles as safeguards that may mitigate the risks that they had identified and place them directly below the relevant risk on the line below. They were also encouraged to use post-it notes for this purpose if they did not think that any of the cards were suitable.
- o **Annotating Safeguards** – as previously the participants were encouraged to record the reasons for selecting certain card(s) as mitigations on post-it notes and place this on the line below.
- o **Challenges of Implementation** – Participants were asked to consider and document what practical elements might challenges the implementation of the safeguards e.g. legal/organisational/social/technical barriers and record these on post-it notes on the line below as previously.
- o **Discussion** – Groups were encouraged to discuss throughout the exercise with this forming the research data considered here. Following the completion of the IA process open summative discussions were held which included: reflection on the IA process, value of the cards as a reflective tool, substantive ethical questions that arose for their technology, impact on their future work.

These workshops were audio recorded, and IA sheets retained for analysis. The audio was anonymously transcribed, and then the researchers conducted detailed thematic analysis of the transcripts. Following best practice in thematic analysis (Braun and Clarke, 2006), initial codes were formed inductively and then through reflexive debate and discussion, these were refined into the key themes presented here.

## Part III – Results and Discussion.

*Table 4 Table of Participants*

| Group Name | General description of participants. |
|---|---|
| **Internet of Things – IoT** | 6 participants 3 male, 3 female. All research or academic roles (namely postdoctoral researchers; assistant, associate and full professors). |
| **Smart Transportation – ST** | 2 participants, 1 male, 1 female. Both academic 6-27 years of experience. |
| **Mixed Reality – MR** | 4 participants 2 female, 2 males, all academic. |
| **Affective Computing - AC** | 4 participants 3 male 1 female all academic. 4-18 years of experience. |
| **Software Development – SD** | 4 participants 3 male 1 female – from financial, software and HR industries.[12] |

---

[12] Due to Covid, unable to get further data at this time, as this data physically held at University, for anonymity purposes. But is possible to add years of experience, if useful.

The following sections will discuss the value of our cards as a tool, their impact on the technology design and how they structure ethical reflection practices.

1. **Cards as a Tool.**

There was flexibility shown in how the cards enabled discussion, within the domain of the IA process. We observed three approaches for how participants selected cards as their ethical risks, which we frame as: ethical clustering, ethical sorting, and ethical replacement. We then consider how the cards had value as anchors for ethical discussion once they were chosen.

**Ethical Clustering:**

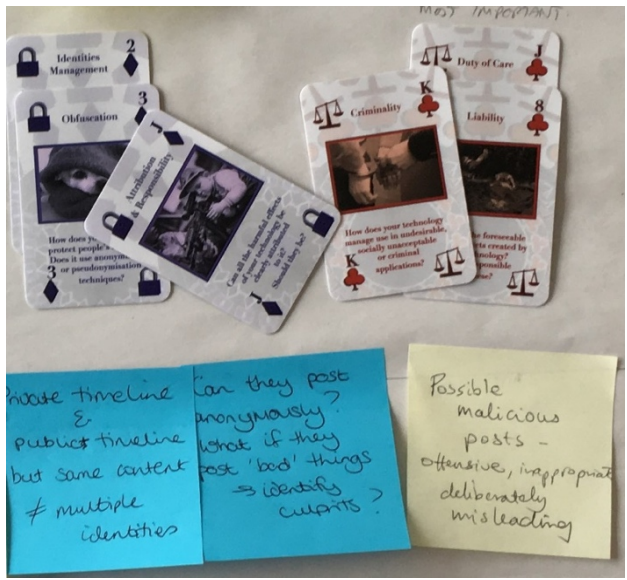

*Figure 4 An ethical cluster from the ST group*

Many participants found sorting the relevant from irrelevant cards initially difficult, as they all could be applied to the technology under consideration to some extent. Utilising the affordance of the cards, we observed some groups doing what we term 'ethical clustering'. Groups would thus cluster the cards together according and construct associated risks as clusters of cards rather than individual principles. The combination of cards enabled more nuanced understandings of different aspects of risk approach through card combinations. We can see this in figure 4 where linked issues around liability also tie into criminality and duties of care; or in thinking about identities management, they recognise the importance of both obfuscation strategies and that there are parties responsible for doing this. The importance of the physical arrangement of the cards is shown here with the Attribution and Responsibility card serving to bridge between the two other clusters of cards.

One participant documented their team's choices and clusters thus far, with card choices including obfuscation, identities management, secrecy, data breach management, and confidentiality as shown in the quote below.

> *"I think most of mine are based on identity because our system requires users to essentially self-identify. So, we've got, for <u>obfuscation</u> how does your technology protect peoples' identities? Does it anonymise or use anything like that? <u>Identities management</u>, does your technology enable systems to hold and manage more identities? <u>Secrecy</u>, does your technology keep secrets? That's kind of the same thing. <u>Data breach management</u>. I guess it is kind of separate. It's more about how securely and everything is stored. And <u>confidentiality</u> is kind of the same thing." Group MR*

Clustering together the cards either physically or discursively as in the example above was also part of the sensemaking process of the workshops. Such an approach demonstrates how despite separating ethical principles onto individual cards, the cards afford the clustering and recombination of principles to do justice to the complexity and variety of the technology under consideration. By placing a group of cards next to each other in a cluster, the complexity of the reflection required by the specific context and technology is made material and visible for others to see, with the cards demonstrating how they enabled nuanced specific discussion of the ethical complexity of an emerging technology.

**Ethical Sorting** - "Are we going by suit or are we going by shuffling and dealing them out? . . .I'd say we shuffle them up."  Group SD

In deciding on their relevant principles associated with their overall ethical risk, the groups commonly divided up the single pack of cards between them. They then sorted their share into relevant and irrelevant individually before discussing as a group which ones should be included amongst their final five. The sorting of the ethical principles was facilitated by the physicality of the cards allowing the user to flick through them quickly and sort into piles or pick out important principles for consideration later. The physical process mitigated some of the initial difficulty expressed by some participants who may have found the range of principles intimidating at first. Breaking their considerations down through addressing the cards they were able to consider the principles one by one, relatively quickly, selecting the ones that they felt most appropriate to anchor their consideration of overall ethical risk.

> *"PA3 - I don't know. I think we've ended up with just too many cards. We've got to narrow these down to five. I guess we just go through what we've got, right?.*
> *PA4 -  Yes, and if you've got any one that's similar, we can choose between them, can't we?"*
> *Group SD*

Once individuals had sorted out their candidate cards, then a process of negotiation followed whereby the group whittled down their selections.  This involved individuals proposing their selected cards for consideration, often by reading out the title ('I've got…X'), the question/principle and then others voicing their opinion about the relevance and where it should be placed. This allowed and encouraged all participants to engage, with each participant volunteering their cards and teaching the others about their content. The discursive process facilitated by the cards was accommodating and *levelled the playing field*, reducing the capacity of individuals to dominate, a theme discussed in part 3.

**Ethical Comparison and Replacement** - "Let me read that one again. I think that probably puts it better than that one." Group SD

We observed some teams adopted a more formal method of deciding on their final selection of cards that we characterise as ethical comparison and replacement.

> *PA4    I've got data security. Does your technology protect from unanticipated disclosures? I was thinking about it'll have all sorts of movements and locations and things like that.*
> *PA2    Yes, location privacy.*
> *PA4    Okay, so maybe that's more specific.*
> *PA3    Because that could be a disaster, couldn't it, if you were sharing your family plan in your insurance and it's like, well, what are you doing in this area?*
> *PA2    We don't ask those kinds of questions.*

> *PA4 So is location, is that one more specific? Is that better?*
> *PA1 I think it's a strong one. I think it's a very strong one. It should be, I think.*

This involved taking cards that were considered as candidates for inclusion and then directly comparing them and selecting one. This was also used for deciding on the order of cards by comparing pairs of cards in turn and deciding which one was most important. This pair based comparison continued until a final ranking was decided upon. This was akin to a physical 'bubble sort' (Astrachan nd) with the cards facilitating the comparison with a pair being held up or pointed to with the 'winner' being placed and moved onto the process board in the position decided resulting in the final selection.

### **Anchors - Appropriate and Inappropriate.**

Participants sought reference points with which to elucidate, explain and demonstrate their points. They drew these reference points from their own experience or from the cards in the workshop and used them to anchor the discussion. Whilst the cards were designed to spark, encourage and structure discussion through their open-ended questions, participants already came to the table with an understanding of the technology, its design and operation. This was now anchored in the ethical principles by the process of sorting and selecting cards. The role of the cards as anchors showed how they were used to complement existing understandings of the system, not reducing or closing off discussion but instead acting as valuable reference points, tying the discussion to principles which made ethical consideration and reflection more tractable, limiting its ability to shift and move**.** We characterise our cards as providing *appropriate anchors* for discussion which were in contrast to other *inappropriate anchors* used by the participants which we discuss below.

How the cards complemented and acted as anchors for existing knowledge can be seen by the material record of the discussion left on the IA process board and for example how one card had the flexibility to be interpreted and act as an anchor by groups for different technologies in different ways. A good example is the 'privacy in public' card (see Figure 5). Where three groups interpreted it, and used it to anchor their discussion, in different ways:

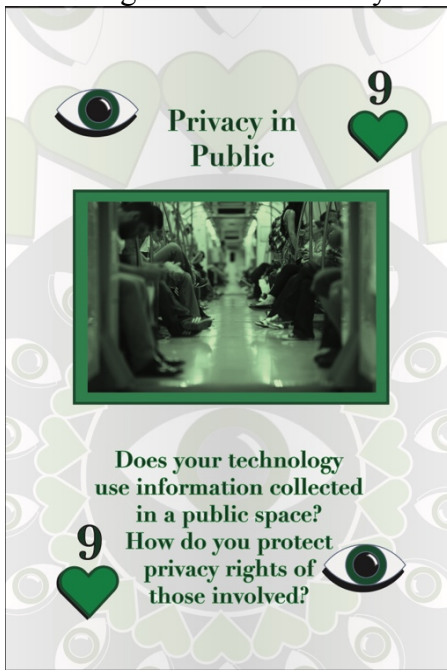

*Figure 5 Privacy in Public Card*

Group IoT used it to anchor their thinking about the ethical dimensions of consent for storing *information collected in public* where photographs, video and audio may capture both families and the general public.

This contrasts with Group S who were talking about tracking for car insurance purposes and were concerned with the data that may be collected in public and how it may be *misused* for other unintended purposes.

Group MR used the card to anchor their discussions about the potentially *public nature of the content* associated with a gift and where and to whom this content may be revealed, potentially inappropriately. All three are pertinent ethical issues, and due to their interpretive flexibility, the cards are able to act as appropriate anchors that enable users to draw connections between ethical and legal principles.

In contrast, there can also be pre-existing *inappropriate anchors*. Here, we saw sensemaking strategies and tacit knowledge shaping ethical rationalising about best practice. This was particularly prominent in reference to large corporations for their good or bad actions e.g. being a bit 'Ben and Jerry' in reference to socially responsible companies or 'don't be evil' in relation (albeit sarcastically) to Google's famous moniker. A good example is from Group AC discussing smart speaker data collection, where we see a blend of two notions of ethical good and bad. Amazon is suggested as an example of poor ethical practice, but the notion of 'Scout's honour' could be positive, perhaps hinting that researchers feel they are held to different moral standards (and more trustworthy). What both examples have in common is that they are both anecdotal reference points based on partial considerations of ethics in different contexts. Generalising lessons from such examples to inform consideration of their technology could be seen as inappropriate reference points (anchors) for ethical reflection.

> *IE2   I'm trying to figure out how Amazon convinced people that Alexa doesn't actually record and is just ephemeral, here and gone. Because really it is sending that whole footprint over, it's not even keeping it there, so, how did they do that?*
>
> *IE1   Maybe they just don't bother. So, we're interested in transparency, Amazon aren't.*
>
> *IE5   So, one of the solutions must be we tell them explicitly, right? So, that is the transparency solution, that doesn't necessarily solve the honest and trust bit, but at least then we've told them we're not storing anything. Honestly. Scout's honour." Group AC*

The use of inappropriate anchors, such as company practice, illustrated how participants sought to navigate the ethical consideration of technology through reference points. We observed that the Moral-IT cards provided a valuable counterpoint to such anecdotal and partial examples and as a tool for discussing ethical dilemmas we argue they can provide flexible, generalisable and *appropriate* anchors for reflection on their moral responsibilities.

This is similar to the observations of Felt, Fochler and Sigl (2018) who argue that cards provide rhetorical resources and a 'narrative infrastructure' to enable individuals to discuss issues that may previously be unfamiliar with. We contend that the Moral-IT cards serve to flexibly structure the discussions that were driven by the knowledge and understanding of the participants. They anchor these disparate interpretations and perspectives in a shared sensemaking experience aided by the cards. This led to improved communication, mutual understanding and a greater grasp on the pragmatic implications of the ethical considerations at hand. When compared to inappropriate anecdotal and partial anchors employed by the participants, the value of this emergent element of the cards is demonstrated. We can begin to understand how, through the cards, discussions are 'anchored' and provide insight into how more abstract principles are understood and employed in practice by technology developers.

## 2. Cards Structuring Ethical Debate

We will now expand on how we observed cards impacting the ethical debates around design of new systems in more depth. We focus here on 3 key observations: 1) the cards *levelled the playing field* between participants in terms of ethical knowledge and engagement with discussions 2) that the cards provide insights into how participants view ethics, something we frame as 'intertwined ethics' 3) How the starting point for ethical discussions is significant and how the cards impact this.

**Levelling the Playing Field on Ethics.**

As we have seen, the Moral-IT cards structure discussion through the sorting, turn taking and negotiation process. We observed that this approach means each member of the group is enabled to participate, regardless of their knowledge, expertise and seniority within the group. The cards ask open questions and treat everyone equally in the face of the question, prompting an answer about how 'your' technology deals with this issue. This requires deliberation both as an individual and a group, encouraging an individual response to be generated which could then be shared with the group to start or continue the discussion. Furthermore, due to the collaborative structure of the impact assessment board, no single person has privileged access all cards and thus they need to understand the other principles through listening to and negotiating with the others who have other cards, as this exchange below referring to a negotiation between the use of the bias and prejudice and autonomy cards, shows.

> *PA4   I feel like the bias is…I still think that's high, very high.*
> *PA2   One group of people differently versus there's the autonomy we've got on top.*
> *PA1   Those are two different ways of looking at the same thing. It's like autonomy is I decide where I'm going or what I'm doing, when I'm doing it. Bias and prejudice is the organisation that's providing my choice, deciding I'm allowed to do things or the groups of people are allowed to do.*
> *PA4   And I feel like that is in ethical terms the worst.*
> *PA1   I think she's trumped us.*
> *PA2   I'm willing to yield. [Unclear].*
> *PA3   Yes.*
>
> *Group SD*

With the support of the cards, we see above that a position of one principle is being advocated as a more important choice, and a participant is able to convince others of their point of view, using card game language of being 'trumped'. Using the cards to support this discussion facilitated shared understanding comparison and ultimately agreement about the choice and priority of a principle.

We also observed the cards made ethical reflection less daunting by deferring responsibility for the judgement and questioning to the card itself and its authority. This enabled participants to put opinions across, not as their own ungrounded view, but instead supported by the card which mandated a response. The cards can empower participants to deliberate on ethics of technology (even if this is new territory for them) by providing valuable, provocative, discursive resources and a structured process of use.

**Intertwined Ethics.**

The cards produced insights into how the ethics of technology was viewed by the participants, something we call 'intertwined ethics'. They were concerned with a wide range of ethical dimensions from managing their own intentions as developers, doing the 'right thing' and managing consequences of their technology being used in both intended and unintended ways e.g. Group S discussed if they make effort to build an ethical ML system, they have limited capacity as author to stop someone appropriating it.[13]

The *intertwined* nature of ethical consideration demonstrated by our participants showed us how they navigate ethics in practice. It highlights that well known ethical perspectives, such as virtue or consequentialist ethics, that are often used to categorise ethical judgements may not be individually satisfactory or appropriate for dealing with the technology development context that the Moral-IT cards are intended to serve. Instead, they are mixed, combined, partially used or even neglected.

Examples of how participants were concerned with the consequences and use of their technology included Group MR concerned about how their mixed reality gift wrapper could be misused by attaching unwelcome digital content (e.g. an embarrassing song)[14]. Group IOT's system enables elderly users to manage their creative content for their own wellbeing, but also to leave a digital legacy after they are gone. The temporality card structured this (and other) discussions around managing future impacts. For Group IOT, they worried about consent and who manages the user's identity after death, where a system might show the user in an unfavourable light identifying a need for stewardship over this. As they state "So, it's that identity, the control over identity and how we make sure that the person is remembered as he or she wanted". Group AC were concerned about how to communicate the ethical nature of their technology and their early 'virtuous' design choices. They were keen to counter what they considered to be misconceptions around how the users would understand their affective computing system.[15]

### *Importance of the Starting Point:*

Of central importance to the intertwined considerations was the ethical starting point of the discussion. As noted above, Group AC shaped their discussion around the communication of the operation of their technology, partly as they acknowledged the perception that *"Computer vision people are not worried about people being creeped out."* They wanted to set themselves apart from their view of research community through their 'virtuous' ethical practice by designing with ethical aspects of user privacy and control in from the start. Their discussion then focused on how to communicate with the user about how the system operates, in contrast to their view of the computer vision community who were not worried about 'creeping people out'. From such a starting point, the challenge was one of communication in addition to the development of the system itself.

A similar starting point was shown in other groups who insisted that they were 'good boys' or perhaps tongue in cheek saying 'Scouts Honour' almost signifying a recognition with these statements that such assurances of good intentions and virtue have to be taken on faith as there was no guarantees when faced with the pragmatic reality of the constraints and expectations in which the technology operates.

---

[13] *"It is that standing on the shoulder of a giant thing a little bit. I might write some software from a completely ethical point of view, what can I do to make sure that people who use my software use it in an ethical way? What I had in mind for it might not be what somebody sees and go, I know what I can use that for. So if I write a nice little machine learning system, somebody can take that and do things that I wouldn't be agreeable to as the author."* GROUP SD

[14] *'it can be that they open it in a public space and something which they didn't intend to reveal, like them singing a song which they might feel they wouldn't want to happen in a public space, it will embarrass them leaking out in public…'* GROUP MR

[15] *"Is it fair then to say that one of our primary risks is actually gaining people's trust/understanding about what this is all about. Just the getting over misconceptions."…"My own version of risk would be people not understanding the system and thinking that we're doing all kinds of mean things with their data. Basically, them not necessarily trusting that we're not using their data."* GROUP AC

Our commercial group SD highlighted that for them, ethics were secondary to legal constraints as the starting point for what a company was allowed and not allowed to do.[16] The priority of the financial bottom line and the law, contrasts with concerns our academic technology developers raise, who are less subject to the practical necessity of making a profitable product and profitable company. This suggests that ethics by design tools need to align with and ideally enhance the business practices of a company. For our group they also raised the fact that not all businesses begin from the same starting point, and indeed may use differences as their competitive advantage. They were concerned of how to resolve the tension that if they, as a responsible company took measures to ensure ethical practice, there would be potential for a disruptive competitor to undercut them by not incurring any cost of producing more ethical technology. As they state "*just because you can do something doesn't mean you should do it, but there will always be somebody who is willing to do it. And that's a worry because they can come in and they have access to exactly the same markets and might not be as scrupulous as we are*".

The range of starting points in the discussions raise the importance of reflexivity in an ethics by design exercise, whether facilitated by the cards or not. Exploring the unspoken embedded assumptions can be key. For example, here developers are for example motivated by demonstrating that they are 'virtuous'. Or being subject to financial business priorities has a significant impact on the way that ethics are discussed and by extension, implemented.

3. **Cards Impacting Technology Design**

In thinking about ethics by design, we conclude by discussing how the cards structured our participants thinking around technical design choices, and we do this with one detailed example from Group AC. They focused on the challenge of communicating how their system worked. They sought appropriate metaphors to demonstrate what data it collected and how this was stored, shared and deleted under user control. Whilst these measures speak to a number of ethical principles represented on the cards such as transparency and user control, the discussion went onto highlight how the method of conveying the operation of the system could be considered to be ethically problematic in itself.

"

*So, to sum that up, I've written visualisation as a challenge of implementation. Which is essentially it, isn't it? And is it a visualisation as something that represents what's happening or is it a view onto actually what is happening, is the challenge there. I don't know if it was a thing or just a design, the USB stick that swelled up as it became full of data, and you plug it in and it's slim because it's empty and as you put files on it, it goes…*

*So, that's clearly not a reality, it's a visual trick.*

---

[16] *"So where I've worked in ethics before we usually start with the law, then we work down to policy and then you work down to the cases covering them. So it's like a hierarchy, it goes…down from law."* Group SD

*So, do we need something like that, water, liquid, or is it somehow you the actual process at work because it's transparent? We're not facilitating this."*

*Group AC*

The group discussed a range of ways of demonstrating what data a camera would record, and how it was held, through metaphors such as filling, emptying and diluting liquid as representative of data. They also sought to 'decamera the camera' and find a way to communicate that the camera would not actually take and store a picture but simply was used to collect certain points of interest necessary for the operation of the system. The discussion soon turned to whether these metaphors were actually helpful, and transparent demonstrations of how the system worked, or simply 'parlour tricks' that misrepresented the operation of the system as the data for example was not liquid and sharing it was not done through diluting it along with other users' data in aggregate form.

The group did not reach a conclusion as to what constituted an appropriate explanation of how a system worked but the discussion raised the pragmatic issue that ethical practice, in this example, attempts to explain a system were hindered by a lack of an appropriate method. Metaphors of demonstration such as filling and emptying may be appropriate to a prior more mechanical system but they were not deemed to be for a computational system. Cards like meaningful transparency can start a discussion, which is valuable, but that does not mean the cards will provide the solution. As a reflective tool, there is still significant value in provoking these discussions and thinking about what it means for technology design. The appropriateness of the language used to discuss it and the expectations this raises and how this may impact its operation also need further attention.

## Part IV - Conclusions

This paper has documented the rationale, development process and substantive detail on a new ethics by design tool: The Moral-IT Cards. It has also provided insights into how this tool works in practice, documenting key empirical findings and insights into doing ethics by design in practice using the cards. We conclude by pulling out some key takeaway findings from the development, evaluation and testing of this design probe.

The cards embed thinking about ethical issues in design, as opposed to moving this to external, outside assessment from ethicists or social scientists. They require technologists to reflect on and take responsibility for their design choices. Furthermore, the IA board enabled structured, rich reflection and proved a valuable tool for making the subject matter more accessible to technologists. We found it interesting that issues such as temporality emerged as a dimension of responsibility for participants, for example in the IOT Group example of how they'd manage legacy data from the system. The discussion of finding mechanisms to demonstrate how a new technological system works (as opposed to how people think it works), was highlighted in the AC Group. This showed that the cards pose questions that enable reflection on topics such as system legibility and appropriate metaphors. However, there remains a lot of work to find strategies that best answer the quandaries posed.

Cards are a valuable medium for communicating complex ideas and structuring practical discussions. We observed the physicality of cards is valuable for enabling ethical clustering; sorting; and comparison, making ethical deliberations material. That physicality can be a weakness too, as once cards are printed, they can be harder to change and it is time consuming to create such a high-fidelity physical prototype. They provide appropriate anchoring points of discussion, and help navigate what may be inappropriate tacit anchors too. Importantly, for group deliberations, they help to *level the playing field* and enable collaborative deliberation on ethics. This corresponds with Felt, Fochler and Sigl (2018) observation of how *"unavoidable hierarchies . . . are, in our experience strongly moderated by the card-based format."*(pp213) or Brey (2017) desire for 'serious moral deliberation under conditions of equality' . We argue collaborative card-based methods can serve to democratise ethical deliberation and discussion processes. We also found that our cards showed how ethical frameworks in design are intertwined in practice, and not neatly separated into distinct forms. On one side, we observed participants acting virtuously by going through the IA process, but also raised concerns about consequences over time and impact of their work. Beyond the practicalities, we received feedback that the medium and aesthetic is viewed as fun. For example, a participant from Group SA stated "*I think it's fascinating these cards because I've been doing software forever and we generally get these things applied retrospectively... It can be you were sitting around with the development team and you're like, okay, we're going to have a session this afternoon, we're going to bring these cards in and before you'd even start on the system you're going over these things and you're building in from the get go. I can just see how great it is.*" In conclusion, the Moral-IT Cards and Board served their purpose of being a valuable design probe for enabling discussions complex value tensions and showed practical potential for doing ethics by design in a technology development environment.

# Appendix 1. TABLE OF CARDS CHOSEN

| Group Name | Group explanation of project | Top Ethical Risk | Risk Cards Chosen (incl. ranking 1-5) | Safeguard Cards Chosen (and what they link to) | Other Cards used and where |
|---|---|---|---|---|---|
| IoT Group 1 | Physical Artefact – Repository for Personal Memories and external content | Exposure of Sensitive Personal Data – (Risk Minimisation Card next to it) | Clustered (NB First workshop we asked them to choose 10 cards)<br><br>• Data Security<br>• Physical Safety<br>• Trustworthiness<br>• Attribution and Responsibility<br>• Duty of Care<br>• Secrecy<br>• Temporality<br>• Identities Management<br>• Obfuscation<br>• Special Categories of Data | •Confidentiality<br><br>•Data Breach Management<br><br>•Resilience and Low Redundancy<br><br>•Wellbeing<br><br>•Taking Responsibilities<br><br>Spectrum of Control Rights | • Legibility and Comprehension (Challenge)<br><br>• User Empowerment and Negotiability (Challenge)<br><br>Autonomy and Agency (Challenge) |
| IoT Group 2 | Physical artefact with cloud services.<br><br>It is a piece of technology to assist capturing personal memories to be blended with facts (external media) | Capacity to Consent<br><br>Changing Status<br><br>Mental Capacity | • Privacy in Public<br>• Liability<br>• Rule of Law<br>• Duty of Care<br>• Trustworthiness<br>• Confidentiality (equal with above)<br>• Temporality<br>• Special Categories of Data<br>• Accessibility<br>• Privacy Virtues | Safeguards written on post its – no cards used | |
| ST | Mobile App for (inter campus) bus users to help practically and entertain and inform along the way | UGC (User Generated Content) – What data is collected and shared publicly | • Cluster 1<br>• Identities Management<br>• Obfuscation<br>• Attribution and Responsibility<br>• Cluster 2<br>• Duty of Care<br>• Liability<br>• Criminality<br>• Cluster 3<br>• Trust<br>• Trustworthiness<br>• Cluster 4<br>• Temporality | | |
| MR | A platform for authoring digital content to be assigned to a physical object. The platform implements access control for sharing | Inappropriate Sharing – e.g Bullying, Illegal Content, extremism | • Wellbeing<br>• Privacy in Public<br>• Limited Data Collection<br>• Obfuscation<br>• Risk Minimisation | Data Security<br><br>Due process | |
| AC | CV (Computer Vision) based tracking of interactions with games, data driven souvenirs and analytics. | Peoples' understanding of how it works and what data it stores | What People Think It Does - CLUSTER 1<br><br>• Spectrum of Control Rights<br>• Temporality<br>• Transparency Rights<br>• Limited Data Collection<br>• Privacy in Public<br>• Legibility and Comprehension<br><br>What System Does – CLUSTER 2<br><br>• Trust<br>• Accessibility<br>• Rule of Law<br>• Confidentiality<br>• Integrity<br>• Obfuscation | | |
| | | | • Criminality<br><br>Agency of User (Input of User) Persistence (taking it away – still meaningful) Persistence of ethics – What is right now might not be in the future<br><br>• User Empowerment and Negotiability<br>• Trustworthiness | • Meaningful Transparency | |
| SD | Insurance – Tailor insurance cost by consumer behaviour (NB discussion focused on car insurance) | Discrimination<br><br>Limit Autonomy | 1. Overt Bias and Prejudice<br>2. Autonomy and Agency<br>3. User Empowerment and Negotiability<br>4. Privacy in Public<br>5. Location Privacy | • Meaningful Transparency<br>• Duty of Care<br>• Consumer Protection<br>• Taking Responsibilities<br>• Limited Data Collection | • Resilience and Low Redundancy (Challenge of Implementation)<br>• Precautionary Principle (Challenge of Implementation)<br>• Make what is perfect more human (annotation to safeguard) |

## Appendix 2. TEXT OF NARRATIVE CARDS.

| Card Title | Card Text |
|---|---|
| State of the Art | Are there any new technical approaches underpinning your technology? Consider these, if they are riskier than current approaches and list 2 reasons why. |
| Safeguards | List two safeguards that address the risks posed by your technology. Also provide two practical constraints to implementing these |
| The Technology | Briefly describe what your technology is and ow it works. |
| Surfacing Risks | List the three biggest risks your technology poses. |
| Stakeholders | List three direct or indirect stakeholders impacted by your technology |
| Use Case | Reflect on two contexts of use for your technology. Describe them below. |